\documentclass[prl,a4paper,showpacs,twocolumn,superscriptaddress,longbibliography]{revtex4-1}

\usepackage{amssymb}
\usepackage{amsmath}
\usepackage{amsfonts}
\usepackage{graphicx}
\usepackage{bm}
\usepackage{color}
\usepackage{multirow}
\usepackage{natbib}
\usepackage{hyperref}
\usepackage[normalem]{ulem}
\usepackage{relsize}

\DeclareMathOperator{\Tr}{Tr}

\DeclareMathOperator{\sgn}{sgn}
\DeclareMathOperator{\re}{Re}

\usepackage[final]{pdfpages}
\usepackage{pgffor}

\makeatletter
\AtBeginDocument{\let\LS@rot\@undefined}
\makeatother

\begin{document}

\title{Mesoscopic Stoner instability in open quantum dots: 
suppression of Coleman-Weinberg mechanism by electron tunneling}

\author{I. S.~Burmistrov}

\affiliation{\hbox{L.~D.~Landau Institute for Theoretical Physics, acad. Semenova av. 1-a, 142432 Chernogolovka, Russia}}

\affiliation{Laboratory for Condensed Matter Physics, National Research University Higher School of Economics, 101000 Moscow, Russia}

\author{Y. Gefen}
\affiliation{Department of Condensed Matter Physics, Weizmann Institute of Science, 76100 Rehovot, Israel}

\author{D. S. Shapiro} 

\affiliation{Department of Physics, National Research University Higher School of Economics, 101000 Moscow, Russia}
\affiliation{Dukhov Research Institute of Automatics (VNIIA),  Moscow 127055, Russia}
\affiliation{\hbox{V. A. Kotel'nikov Institute of Radio Engineering and Electronics, Russian Academy of Sciences, Moscow 125009, Russia}} 

\author{A. Shnirman}
\affiliation{\mbox{Institut f\"ur Theorie der Kondensierten Materie, Karlsruhe Institute of Technology, 76128 Karlsruhe, Germany}} 

\affiliation{Institut f\"ur Nanotechnologie, Karlsruhe Institute of Technology, 76021 Karlsruhe, Germany}


\begin{abstract}
The mesoscopic Stoner instability is an intriguing manifestation of symmetry breaking in isolated metallic quantum dots, underlined by the competition between single-particle energy and
Heisenberg exchange interaction. Here we study this phenomenon in the presence of tunnel coupling to a reservoir. We analyze the spin susceptibility of electrons on the quantum dot for different values of couplings and temperature. Our results indicate the existence of a \emph{quantum phase transition} at a critical value of the tunneling coupling, which is determined by the Stoner-enhanced exchange interaction. This quantum phase transition is a manifestation of the suppression of the Coleman-Weinberg mechanism of symmetry breaking, induced by coupling to the reservoir. 
\end{abstract}

\maketitle

The physics of quantum dots (QDs) has been the focus of theoretical and experimental  study for three decades~\cite{Alhassid2000,Wiel,ABG,Hanson,Ullmo2008}. A major breakthrough in this field  was the introduction of the so-called  ``universal'' Hamiltonian \cite{KAA}, rendering QDs  as zero-dimensional objects. This is valid for  metallic QDs, characterized by the Thouless energy being larger than the mean single particle level spacing, $E_{\rm Th}\gg \delta$. The universal Hamiltonian comprises a charging energy term which leads to  Coulomb blockade \cite{CBa,CBb,CBc,KamenevGefen1996,SeldmayrLY}. 
An additional term in the  universal Hamiltonian is a  ferromagnetic Heisenberg exchange term. Even relatively weak exchange interaction, $J\lesssim \delta/2$, seems to be  important for a quantitative description of transport experiments in QDs at low temperatures, $T\lesssim \delta$ \cite{QDLowTa,QDLowTb,QDLowTc,QDLowTd}.  Moderate exchange, $\delta/2 \lesssim J<\delta$, \cite{footnote0} gives rise to \emph{mesoscopic Stoner instability}:
the emergence of a finite (but non-extensive) value of the total electron spin, $S$, in the ground state of an isolated QD \cite{KAA}. In the vicinity of the transition, $\delta-J\ll \delta$, the ground-state spin is estimated as $S = J_*/(2\delta) \gg 1$, where $J_* = J\delta/(\delta-J)$ denotes the Stoner-enhanced exchange interaction. 
At $J=\delta$ 
an extensive part of electron spins becomes polarized, i.e. a Stoner phase transition to a macroscopic ferromagnetic phase takes place. A non-zero value of $S$ gives rise to a finite Curie spin susceptibility at low $T$ \cite{KAA,BGK1,BGK2,Saha2012}. Spin-charge coupling leads to  signatures of the mesoscopic Stoner instability in electron transport through QDs \cite{KiselevGefen,BGK1,BGK2,Koenig2012}.

The physics of the mesoscopic Stoner instability in an isolated QD is marked by total spin conservation. It is an example \cite{Saha2012}  of the Coleman-Weinberg mechanism 
for the emergence of spontaneous symmetry breaking \cite{ColemanWeinberg}.
Does the Colemann-Weinberg mechanism survive electron tunneling dynamics between the QD and the reservoir? Addressing this question is not straightforward, given the fact that spin conservation is then broken, resulting  in a nontrivial dissipative dynamics of $S$ \cite{Shnirman-PRL,Shnirman-JETP}. Similarly to the problems of a localized spin in an electronic environment \cite{PhysRevB.73.212501,Tretyakov,PhysRevB.85.115440} or that of an itinerant
magnetization \cite{ChudnovskiyPRL,BaskoVavilovPRB2009}, the 
equation of motion for the total spin on the QD assumes the form of the Landau-Lifshitz-Gilbert-Langevin (LLGL) equation.  
We note in passing  that  in Refs. \cite{Shnirman-PRL,Shnirman-JETP} the LLGL equation has been derived under the assumption that the tunneling between the QD and reservoir does not change the value of
 $S$.  

\begin{figure}[t]
\centerline{\includegraphics[width=0.5\textwidth]{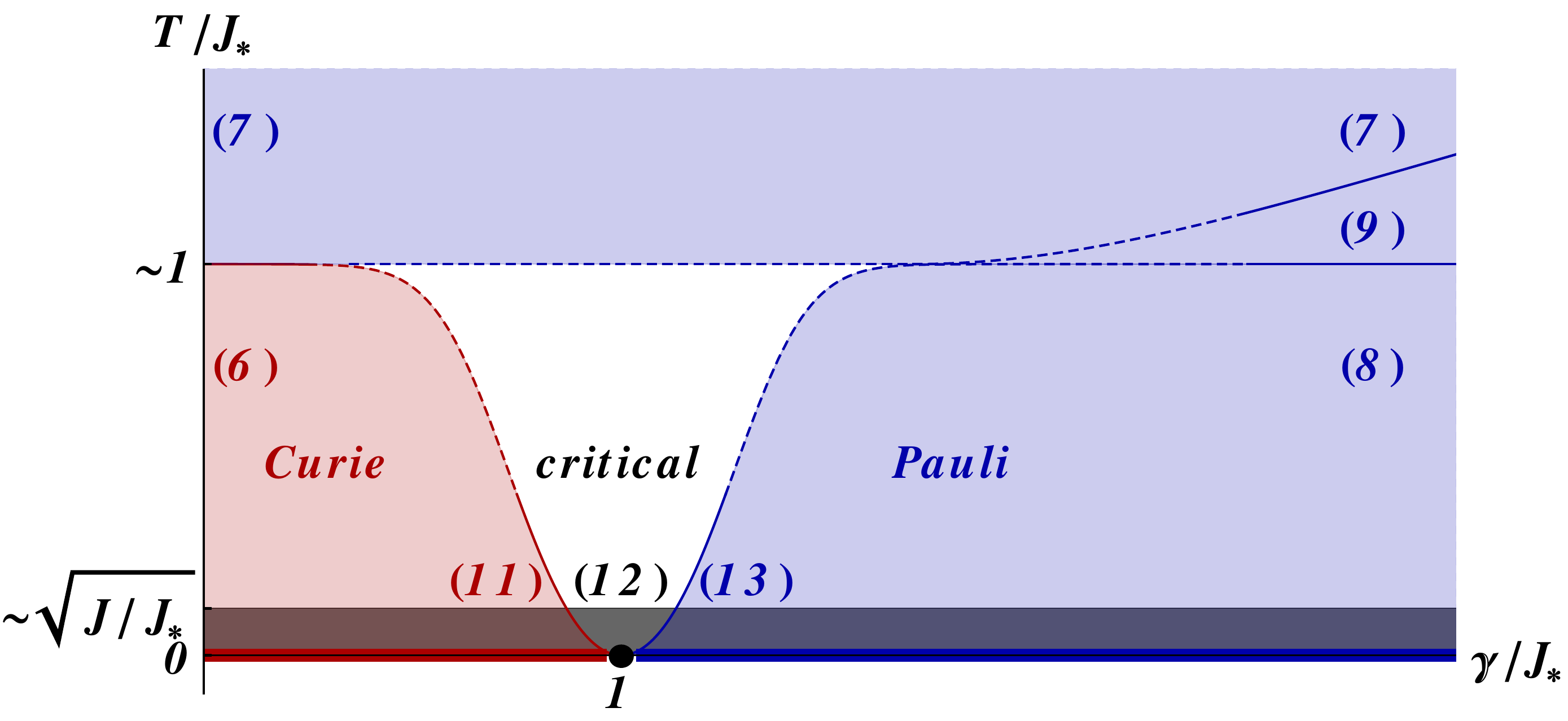}}
	\caption{(Color online) A sketch of the phase diagram for the case $\delta-J\ll \delta$. The red color indicates the region with a Curie-type spin susceptibility above the zero-temperature ordered phase. The blue color indicates the regions with the Pauli-type spin susceptibility. The black dot indicates the position of the QCP. The thick solid red (blue) line corresponds to the zero-temperature ordered (disordered) phase. The thin solid curves corresponds to the crossovers discussed in the text. The dashed lines are guides for the eye. A latin number indicates the equation for the corresponding region of the phase diagram. Black shaded region at the bottom marks the region above which our theory is applicable.}
	\label{fig1}
\end{figure}

The focus of this Letter is the mesoscopic Stoner physics in open quantum dots. We study how tunneling to the reservoir (assigning a broadening $\gamma$ to the single-particle levels) affects the mesoscopic Stoner instability. Addressing the vicinity of the transition to the macroscopic Stoner phase, $\delta-J\ll \delta$, our analysis indicates the existence of the \emph{quantum phase transition} (QPT) at a critical broadening strength, $\gamma_c \simeq J_*$ (see Fig. \ref{fig1}). The quantum critical point (QCP) separates the ordered ($\gamma<\gamma_c$) and the disordered ($\gamma>\gamma_c$) phases. The QPT occurs since tunneling to the reservoir modifies the Coleman-Weinberg (CW) potential, and suppresses the spontaneous symmetry breaking at  $\gamma>\gamma_c$. Our analysis relies on the study of  the spin susceptibility, $\chi$,  of the electrons on the QD.

\textsf{Model.}\ --- A metallic QD tunnel coupled to a reservoir is described by the following Hamiltonian: $H=H_{\rm d} + H_{\rm r} + H_{\rm t}$. 
Here $H_{\rm d}=H_0 +H_{\rm s}$ \cite{KAA}, where 
$H_0= \sum_{\alpha,\sigma} \epsilon_{\alpha} d^\dag_{\alpha,\sigma} d_{\alpha,\sigma}$ is  the free electron part  and  $H_{\rm s} = -J \bm{S}^2$ takes into account the exchange interaction on QD \cite{footnote4}. The free electrons in the reservoir are governed by the $H_{\rm r}=\sum_{k,\sigma} \epsilon_{k}a^\dagger_{k,\sigma}a_{k,\sigma}$. The Hamiltonian $H_{\rm t}=\sum_{k,\alpha,\sigma}t_{k\alpha} a_{k,\sigma}^\dagger d_{\alpha,\sigma}+{\rm
    h.c.}$ describes a multi-channel tunneling junction between the QD and the reservoir with a small dimensionless  (in units $e^2/h$)  tunneling conductance of each channel. The total dimensionless tunneling conductance of the junction, $g$, is assumed large. This assumption allows us to neglect the Coulomb blockade effects associated with the charging energy term in the ``universal'' Hamiltonian \cite{KAA}.
 Here $\epsilon_{\alpha}$, $\epsilon_k$ denote the energies of single particle levels on the QD and in the reservoir, respectively, counted from the chemical potential. The operators $d^\dag_{\alpha,\sigma}$, $a^\dag_{k,\sigma}$ ($d_{\alpha,\sigma}$, $a_{k,\sigma}$) create (annihilate) an electron on the QD and the reservoir, respectively. $\bm{S}=\sum_{\alpha\sigma\sigma^\prime}  d^\dag_{\alpha,\sigma} \bm{\sigma}_{\sigma\sigma^\prime} d_{\alpha,\sigma^\prime}/2$ stands for the operator of the total electron spin in a QD. The vector $\bm{\sigma}=\{\sigma_{x},\sigma_y,\sigma_z\}$ comprises the three Pauli matrices.

In order to address $H_{\rm s}$ we employ the Hubbard-Stratonovich transformation, introducing the bosonic vector field $\bm{\Phi}$. Integrating out fermions, we obtain an effective action in the imaginary time: 
\begin{equation}
S = \frac{1}{4J}\int_0^{\beta} d\tau\ \bm{\Phi}^2 - 
\Tr \ln \Bigl ( -\partial_\tau -\hat \epsilon + \frac{1}{2}\bm{\sigma \Phi}- \hat \Sigma \Bigr ) .
\label{eq:action}
\end{equation}
Here $\beta=1/T$, $\hat \epsilon_{\alpha \alpha^\prime} = \epsilon_\alpha \delta_{\alpha \alpha^\prime}$ and $\hat\Sigma_{\alpha\alpha^\prime} = \sum_k  t^*_{\alpha k} (-\partial_\tau -\epsilon_k)^{-1} t_{k\alpha^\prime}$ is the self-energy induced by the tunneling to the reservoir. In what follows we neglect the mesoscopic fluctuations in the tunneling amplitudes, $t_{k\alpha}$, and approximate the self-energy as  
$\hat\Sigma_{\alpha\alpha^\prime}(i\varepsilon_n)  = - i (\gamma/\pi) \sgn\varepsilon_n \delta_{\alpha\alpha^\prime}$. Here $\varepsilon_n = \pi T(2n+1)$; and $\pi^2 \sum_k |t_{k\alpha}|^2 \delta(\epsilon_k) \to \gamma$ characterizes the uniform broadening of a single-particle level on the QD~\cite{Footnote2}.  It is related to the tunneling conductance of the junction through $g= 4\gamma/\delta$. The spin susceptibility of electrons on the QD  can be computed as \cite{Saha2012}
\begin{equation}
\chi = \frac{T}{12J^2} \Bigl\langle \Bigl| \int_0^\beta d\tau \bm{\Phi}\Bigr|^2\Bigr\rangle - \frac{1}{2J} ,
\label{eq:chi:def}
\end{equation} 
where  the averaging is carried out with respect to the action \eqref{eq:action}.

\textsf{Wei-Norman-Kolokolov trick.}\ --- In order to proceed further one needs to be able to compute the $\Tr \ln$ in the action \eqref{eq:action}. A solution of this complicated problem requires the knowledge of the matrix  
$U(\tau) = \mathcal{T}_\tau \exp [{ \int_0^{\tau}d\tau^\prime\bm{\sigma}\bm{\Phi}(\tau^\prime)/2}]$, where $
\mathcal{T}_\tau$ denotes the time ordering along the imaginary time contour. For an arbitrary trajectory, $\bm{\Phi}(\tau)$, direct evaluation of $U(\tau)$ is impossible. It is possible, though, to perform a transformation  in the functional integral  from the variables  $\bm{\Phi}$ to new variables $\rho$, $\kappa$, and $\tilde{\kappa}$ \cite{WeiNorman,Kolokolov1,Kolokolov2,Kolokolov3a,Kolokolov3b,Kolokolov4}: $\Phi_z = \rho - 2 \kappa \tilde{\kappa}$, $\Phi_- = \tilde{\kappa}$,  and $\Phi_+ = \partial_\tau \kappa+\kappa\rho  -\kappa^2\tilde{\kappa}$,  where $\Phi_\pm  =(\Phi_x\pm i\Phi_y)/2$. 
The Jacobian of this transformation is equal to $\exp(\beta h)$ where $h= T \int_0^\beta d\tau \rho(\tau)/2$ is a one half of the zeroth Matsubara harmonics of $\rho(\tau)$ \cite{Kolokolov4}. This transformation is supplemented by the initial condition $\kappa(0)=0$ which guarantees $U(0)=1$. The $2\!\times\! 2$ matrix $U(\tau)$ can be written explicitly in terms of new variables $\rho$, $\kappa$, and $\tilde{\kappa}$ \cite{footnote3}.

\textsf{Coleman-Weinberg potential.}\ --- As is known from studies of the mesoscopic Stoner phase in an isolated QD \cite{BGK2,Saha2012}, the zeroth Matsubara harmonics of $\rho(\tau)$ 
plays the role of an order parameter. Therefore, our strategy is to derive the effective free energy for $h$
by integrating out the fluctuations with nonzero Matsubara frequency components in the action \eqref{eq:action}. We thus split the field $\rho$ as $\rho(\tau)=2 h+\delta \rho(\tau)$, and integrate over 
$\delta \rho$, $\kappa$, and $\tilde{\kappa}$ within the Gaussian approximation. We then obtain the following free energy (CW potential)  (see Supplemental Material for details \cite{SM}):
\begin{equation}
F(h) = \frac{h^2}{J_*} - h + 2T \re \ln \frac{\Gamma\bigl (1+\frac{ih}{\pi T}+\frac{\gamma}{\pi^2 T}\bigr )}{\Gamma\bigl (1+\frac{ih}{\pi T}\bigr )\Gamma\bigl (1+\frac{\gamma}{\pi^2 T}\bigr )} .
\label{eq:CWpot}
\end{equation}
Here  $\Gamma(z)$ is the Gamma function. 
The origin of different terms in the expression for $F(h)$ is the following. The first term on the r.h.s. of Eq. \eqref{eq:CWpot} is the sum of two contributions, $h^2/J$ and $-h^2/\delta$. The former comes from the first term in the r.h.s. of Eq. \eqref{eq:action} whereas the latter is a paramagnetic part of the thermodynamic potential of free electrons in the presence of a constant magnetic field $2h$. The second term on the r.h.s. of Eq. \eqref{eq:CWpot} appears from the Jacobian of the Wei-Norman-Kolokolov transformation. The 
third term  is the result of integration over dynamical fluctuations of $\kappa$ and $\tilde{\kappa}$ which are coupled to $h$ in the presence of nonzero tunneling. As can be seen from Eq. \eqref{eq:CWpot}, tunneling to the reservoir indeed modifies the form of the CW potential. 
The Gaussian approximation for integration over dynamical fluctuations is justified under the conditions \cite{SM}
\begin{equation}
 |h| , T \gg \max\bigl \{J, \min\{J_*,\sqrt{J \gamma}\}\bigr \} .
\label{eq:condition} 
\end{equation}
Instead of working with the full action \eqref{eq:action}, we can now use $F(h)$ 
for the purpose of analyzing the spin susceptibility. Under conditions \eqref{eq:condition}, the expression \eqref{eq:chi:def} can be simplified to 
\begin{equation}
\chi = \frac{1}{3 T J^2}
\int\limits_{-\infty}^\infty dh\, h^2 e^{-\beta F(h)}\Bigl /\int\limits_{-\infty}^\infty dh \, e^{-\beta F(h)} .
\label{eq:chi:h}
\end{equation}

\textsf{An isolated QD.}\ --- Before turning to the analysis of an open system, it is instructive to recover the CW potential  \eqref{eq:CWpot} for the case of an isolated QD. For $\gamma=0$ $F(h)$ possesses a minimum at $h=J_*/2$. 
At low temperatures, $T\ll J_*$, this minimum is narrow and 
Eqs. \eqref{eq:CWpot} and \eqref{eq:chi:h} yield the Curie law for the spin susceptibility: $\chi = J_*^2/(12 T J^2)$ \cite{KAA,BGK1,BGK2}. At high temperatures, $T\gg J_*$, the minimum at $h=J_*/2$ becomes shallower. The thermal fluctuations then determine the typical value of $h \sim \sqrt{T J_*}$. 
Eqs. \eqref{eq:CWpot} and \eqref{eq:chi:h} 
reproduce correctly the Pauli-type spin susceptibility, known from the exact solution \cite{KAA,BGK1,BGK2}. 
We find from the CW potential \eqref{eq:CWpot} that 
$\chi = c J_*/J^2$ with $c=1/6$. The exact solution, however, yields
the value of $c=1/2$. Such a discrepancy in the prefactor arises since the free energy \eqref{eq:CWpot} reproduces the 
Gibbs weight $\exp(-\beta F(h))$ upto a multiplicative prefactor.

\textsf{Weak tunneling regime, $\gamma\ll J_*$.}\ --- We next analyze the spin susceptibility in the regime of weak tunneling,
$\gamma\ll J_*$. Then the situation is similar to the case of an isolated QD. The free energy $F(h)$ has its minimum at 
$h = J_* [1- 4\gamma/(\pi^2 J_*)]/2$. At $T\ll J_*$ this minimum is narrow and Eq. \eqref{eq:chi:h} yields the Curie law:
\begin{equation}
\chi \sim (J_*/J)^2 \bigl [1- 8 \gamma/(\pi^2 J_*)\bigr]/T .
\label{eq:spin-regI}
\end{equation}
At $T\gg J_*$ the CW potential $F(h)$ has the shallow minimum at $h = J_*[1+\psi^{\prime\prime}(1)\gamma J_*/(\pi^4 T^2)]/2$. Here $\psi(z)$ denotes the di-gamma function. Then the typical value of $h$ is dominated by the thermal fluctuations which are of the order of $\sqrt{T J_*[1+\psi^{\prime\prime}(1)\gamma J_*/(\pi^4 T^2)]}$. Hence at $T\gg J_*$ we find the Pauli-type spin susceptibility:
\begin{equation}
\chi \sim J_*\bigl [1+\psi^{\prime\prime}(1)\gamma J_*/(\pi^4 T^2)\bigr ]/J^2 .
\label{eq:spin-regII}
\end{equation}

Therefore, in the weak tunneling regime, $\gamma\ll J_*$, 
the dependence of the spin susceptibility on temperature is qualitatively the same as in the case of an isolated QD.
\color{black}

\textsf{Strong tunneling regime, $\gamma\gg J_*$.}\ ---
For a strong tunneling, $\gamma\gg J_*$, the CW potential \eqref{eq:CWpot} has the minimum whose position depends on temperature. At low temperatures, $T\ll J_*$, the minimum of $F(h)$ is at $h=0$. The spin susceptibility then is determined by the thermal fluctuations of $h$, which are of the order of $\sqrt{T J_*(1+J_*/\gamma)}$. Thus, for $T\ll J_*$, we find  
\begin{equation}
\chi \sim  J_* \bigl (1+ J_*/\gamma\bigr )/ J^2 .
\label{eq:spin-regIII}
\end{equation}
For intermediate temperatures, $J_*\ll T\ll\sqrt{J_*\gamma}$, 
the free energy \eqref{eq:CWpot} has a shallow minimum at $h = J_*[1-J_*/(6 T)]$. $h$ is typically of the order of $\sqrt{T[1-J_*/(6 T)]}$. Then the spin susceptibility is given by 
\begin{equation}
\chi \sim J_*\bigl [1- J_*/(6 T)\bigr ]/ J^2 . 
\label{eq:spin-regIV} 
\end{equation}
Finally, at $T\gg \sqrt{J_*\gamma}$ the behavior of the CW potential is similar to the one for weak tunneling and high temperatures, $T\gg J_*$. It follows that the spin susceptibility at $T\gg \sqrt{J_*\gamma}$ 
is given by Eq. \eqref{eq:spin-regII}. 

\textsf{Quantum phase transition.}\ --- The above analysis  demonstrates  that at low temperatures, $T\ll J_*$, the minimum of $F(h)$ at non-zero value of $h$ 
survives at weak tunneling, $\gamma\ll J_*$, but disappears at   
strong tunneling, $\gamma\gg J_*$. This suggests the existence of the QPT at $\gamma = \gamma_c\sim J_*$. At $\gamma<\gamma_c$ there is a broken symmetry phase with a non-zero order parameter $\Delta=\lim\limits_{T\to 0} T \chi$. For $\gamma>\gamma_c$ the symmetry is restored such that $\Delta=0$.

In order to further substantiate the existence of a QPT we now consider the low temperature regime, $T\ll \gamma \sim J_*$. One can show that, pushing towards the vicinity of the QCP, the relevant values of $h$ lie within the range $T\ll h\ll \gamma$.
Taking the limit $h,\gamma\gg T$ in Eq. \eqref{eq:CWpot} and then expanding in $h/\gamma$ to the fourth order, we obtain \color{black}
\begin{equation}
F(h) \simeq  (1/J_*-1/\gamma)h^2  + \pi^2h^4/(6\gamma^3)\, .
\label{eq:CWpot:QCP}
\end{equation}

Taking this expression for $F(h)$ literally at  $T=0$ may suggest that there is indeed a QCP at $\gamma_c=J_*$. We recall, though, that setting the temperature to zero is not allowed in view of the inequality  \eqref{eq:condition}. Our strategy to detect the presence of the QCP will be to sweep $\gamma$ near $\gamma_c\simeq J_*$ at the lowest possible temperature,  $T\simeq \sqrt{JJ_*}$. 
We note that Eq. \eqref{eq:CWpot:QCP} resembles the standard form of the Landau free energy with $h$ playing the role of the order parameter. We stress, though, that unlike the Landau free energy which is valid only for small values of the order parameter, here Eq. \eqref{eq:CWpot:QCP} is valid for the entire interval $T\simeq \sqrt{JJ_*} \ll h\ll J_*$.    

The form \eqref{eq:CWpot:QCP} of the CW potential implies a scaling form of the spin susceptibility
$\chi = \sqrt{J_*^{3}/T} f(T_X/T)/J^2$, with a  characteristic temperature scale $T_X=J_*\alpha^2$, and  $\alpha=\gamma_c/\gamma-1$. Notwithstanding the fact that we cannot determine the precise form of the scaling function $f(X)$, as we know $\exp(-\beta F(h))$  only with exponential accuracy,  Eq. \eqref{eq:CWpot:QCP} suffices for the evaluation of the asymptotic behavior of $ f(X)$.

For $\gamma<\gamma_c$ the free energy \eqref{eq:CWpot:QCP} has its minimum at $h= J_* \sqrt{3\alpha}/\pi$. Then, at sufficiently low temperatures and away from the QCP, $T\ll T_X$, we can treat the thermal fluctuations around the minimum as being weak. We then find
\begin{equation}
\chi \sim {J_*^2 \alpha}/{(T J^2)}, \quad T\ll T_X . 
\label{eq:order}
\end{equation}
At high temperatures, $J_*\gg T\gg T_X$, the typical value of $h$ due to the thermal fluctuations is dictated by the quartic term in Eq. \eqref{eq:CWpot:QCP}: $h\sim (T J_*^3)^{1/4}$. Since this value of $h$ is within the range  $T\ll h \ll J_*$, the use of Eq. \eqref{eq:CWpot:QCP} is justified. Using Eq. \eqref{eq:chi:h}, we obtain: 
\color{black}
\begin{equation}
\chi \sim
{J_*^{3/2}}/{(J^2 T^{1/2})} , \qquad T_X \ll T \ll J_*
\label{eq:fluct}
\end{equation}
\color{black}

For $\gamma>\gamma_c$ the free energy \eqref{eq:CWpot:QCP} has a minimum at $h=0$. Then, at low enough temperatures, $T\ll T_X$, and away from the quantum critical point, the quadratic term dominates over the fourth order term in Eq. \eqref{eq:CWpot:QCP}. Thus, the typical value of $h$ due to the thermal fluctuations is given by $h \sim \sqrt{T J_*/\alpha}$. Hence,  the spin susceptibility reads
\begin{equation}
\chi \sim{J_*}/{(3J^2|\alpha|)} , \qquad T\ll T_X .
\label{eq:disorder}
\end{equation}
At  higher temperatures, $J_*\gg T\gg T_X$, the spin susceptibility is given by Eq. \eqref{eq:fluct}.

For $\gamma<\gamma_c$, cf. Eq. \eqref{eq:order}, the spin susceptibility exhibits the Curie-type behavior at $T\ll T_X$, with the effective spin $ \propto J_*\sqrt{\alpha}/J$. The latter decreases as the QCP is approached. For 
$\gamma>\gamma_c$, cf. Eq. \eqref{eq:disorder}, the spin susceptibility at $T\ll T_X$ has the Pauli form with the effective exchange $\propto J_*/|\alpha|$ diverging at the QCP. At high temperatures $T\gg T_X$, cf. Eq. \eqref{eq:fluct}, the spin susceptibility has a critical behavior, $\chi \propto 1/\sqrt{T}$, which is neither Curie- nor Pauli-like. Thus the overall behavior of the spin susceptibility at low temperature is typical for the vicinity of a QCP  (see Fig. \ref{fig1}). 

Since the range of validity of our analysis is limited from below by the temperature $T\simeq\sqrt{JJ_*}$, we can determine the position of the QCP only with a limited accuracy: $\gamma_c=J_* \bigl [1+O\bigl ((J/J_*)^{1/4}\bigr )\bigr ]$. This indicates that our theory becomes asymptotically exact as the system is approaching the bulk Stoner transition at $J=\delta$.

\color{black}

\textsf{Discussion.}\ --- In Ref. \cite{Shnirman-PRL} it has been demonstrated that electron tunneling between the QD and the reservoir  in the regime of mesoscopic Stoner regime induces a Gilbert damping term $g/(4\pi S)$ in the LLGL equation. Our present results imply that the LLGL equation of Ref. \cite{Shnirman-PRL} applies to not-too-large values of the conductance,  $g\lesssim g_c = 8 S$. We note that the QCP corresponds to a value of the Gilbert damping of the order unity. 

Recalling the mesoscopic Stoner phase for  an isolated QD, it is marked by a non-zero value of the total spin in the ground state. This is the case for a finite interval of $J<\delta$. 
A state with a given value of the total spin $S$ is separated by  QPTs (at $J=\delta (2S\pm 1)/(2S+1 \pm 1)$) from states with spin $S\pm 1$. One important implication of our analysis is that the presence of a very weak tunneling, $\gamma\ll \delta$, does not destroy these transitions. We expect that the lines of these QPTs in the $J/\delta$, $\gamma/\delta$ parameter space terminate at  $\gamma \sim \delta$ \cite{SM}. 

The ``universal'' Hamiltonian involves also a term with a Cooper channel interaction. This term represents  superconducting correlations in the QDs \cite{SCinQDa,SCinQDb,SCinQDc,SCinQDd,SCinQDe,SCinQDf,SCinQDg}. Throughout our analysis we have assumed the absence of bare attraction, hence we have disregarded this  Cooper channel interaction.  Moreover, we have also neglected the effect of fluctuations in the matrix elements
of the interaction \cite{Altshuler1997,Mirlin1997}. These corrections are typically
small in the regime
 $\delta/E_{\rm Th}\ll 1$ but may still be responsible for interesting physics beyond the ``universal'' Hamiltonian paradigm \cite{Ullmo2008}.

Another effect we have not considered here  is the fluctuations of single-particle levels on the QD.  Such fluctuations are particularly important in the case of Ising exchange interaction. For the latter, assuming equidistant quasiparticle spectrum, the phenomenon of the mesoscopic Stoner instability is completely absent  \cite{KAA}. The ``universal'' Hamiltonian with an Ising exchange is realizable  in the limit of a strong spin-orbit coupling \cite{AlhassidSO,SOinQDa,SOinQDb,SOinQDc,AF2001}. Considering an Ising exchange and an equidistant single-particle spectrum, the electron spin susceptibility is Pauli-like for all temperatures \cite{KiselevGefen,Boaz}. Accounting  for single-particle level fluctuations (e.g., due to the presence of static disorder in the QD), a mesoscopic Stoner phase does exist for an isolated dot, with an  averaged spin susceptibility yielding a  Curie-type behavior at low temperatures \cite{KAA,LSB,SLB}. In this case, one might expect the emergence of a QPT at a certain value of level broadening (tunnel coupling to external reservoirs), similar to the case of Heisenberg exchange studied here. 

Finally, our results are amenable to experimental verification, employing a single electron box based on nanoparticles made up of materials with parameters close to the Stoner instability. There is a host  of such nearly ferromagnetic materials \cite{Exp1d,Exp1e,Exp1a,Exp1b,Exp1f,Exp1g,Exp1j,Exp1h}.  Promising candidates are the compounds YFe$_2$Zn$_{20}$ ($J=0.88 \delta$) and LuFe$_2$Zn$_{20}$ ($J=0.89 \delta$) \cite{Exp2a,Exp2b}.

\textsf{Summary.}\ ---  
We have studied here the mesoscopic Stoner instability in open QDs, coupled to external fermionic reservoirs. We have developed a detailed theory for  the regime close to the macroscopic Stoner instability, $0<\delta-J\ll \delta$. 
The resulting temperature dependence of $\chi$ suggests the existence of
a QPT at a critical value of the tunneling broadening, $\gamma_c = J_*$. This transition as function of the tunnel coupling strength is between the symmetry broken phase with non-zero value of the total spin in the ground state and spin-symmetric phase. The smoking gun evidence for the QPT is the electron spin susceptibility, switching between Curie and Pauli behaviors.  This QPT (and the onset of the symmetry-conserved phase) marks  the suppression of the Coleman-Weinberg mechanism of symmetry breaking by tunnel coupling  to the reservoir.

\textsf{Acknowledgements.}\ --- We thank I. Kolokolov for very useful discussions. Hospitality by Tel Aviv University, the Weizmann Institute of Science, the Landau Institute for Theoretical Physics, and the Karlsruhe Institute of Technology is gratefully acknowledged. The work was partially supported by 
the programs of Ministry of Science and Higher Education (Russia), the Alexander von Humboldt Foundation, the Israel Science Foundation, the Minerva Foundation, and DFG Research Grant SH 81/3-1.




\section*{Supplementary material}

# ONLINE SUPPORTING INFORMATION

# Mesoscopic Stoner instability in open quantum dots: suppression of Coleman-Weinberg mechanism by electron tunneling

I. S. Burmistrov,[1,2] Y. Gefen,[3] D. S. Shapiro,[4,5,6] and A. Shnirman[7,8]

[1]*L. D. Landau Institute for Theoretical Physics, acad. Semenova av. 1-a, 142432 Chernogolovka, Russia*
[2]*Laboratory for Condensed Matter Physics, National Research University Higher School of Economics, 101000 Moscow, Russia*
[3]*Department of Condensed Matter Physics, Weizmann Institute of Science, 76100 Rehovot, Israel*
[4]*Department of Physics, National Research University Higher School of Economics, 101000 Moscow, Russia*
[5]*N. L. Dukhov All-Russia Research Institute of Automatics, Moscow, Russia*
[6]*V. A. Kotel'nikov Institute of Radio Engineering and Electronics, Russian Academy of Sciences, Moscow, Russia*
[7]*Institut für Theorie der Kondensierten Materie, Karlsruhe Institute of Technology, 76128 Karlsruhe, Germany*
[8]*Institut für Nanotechnologie, Karlsruhe Institute of Technology, 76021 Karlsruhe, Germany*
(Dated: October 22, 2019, v.4)

In this Supplementary Material we present (i) the derivation of the Coleman-Weinberg potential [Eq. (3) of the main text], (ii) the derivation of the applicability conditions for the Gaussian approximation [Eq. (4) of the main text], (iii) the analysis of the effect of tunneling on the first order quantum phase transitions between the ground states with different values of the total spin near and away from the vicinity of the macroscopic Stoner transition.

## S.I. DERIVATION OF THE COLEMAN-WEINBERG POTENTIAL

### A. Transformation of the action

We start from the action defined in Eq. (1) of the main text. The partition function is given as

$$Z = \int D[\boldsymbol{\Phi}]\, e^{-S}, \qquad S = \frac{1}{4J}\int_0^\beta d\tau\, \boldsymbol{\Phi}^2 - \mathrm{Tr}\ln\!\left(G_{\boldsymbol{\Phi}}^{-1} - \hat{\Sigma}\right), \qquad G_{\boldsymbol{\Phi}}^{-1} = -\partial_\tau - \hat{\epsilon} + \frac{1}{2}\boldsymbol{\sigma}\boldsymbol{\Phi}. \tag{S1}$$

Let us introduce the $2\times 2$ matrix

$$U(\tau) = \mathcal{T}_\tau \exp\!\left[\frac{1}{2}\int_0^\tau \boldsymbol{\sigma}\boldsymbol{\Phi}(\tau')d\tau'\right]. \tag{S2}$$

We note that this matrix has the unit determinant, $\det U(\tau) = 1$, and satisfies the initial condition $U(0) = 1$. In particular, at $\tau = \beta$ it can be cast into diagonal form as

$$U(\beta) = V^{-1} e^{\beta H \sigma_z} V \tag{S3}$$

with proper matrix $V$. It is convenient to make time-dependent rotation under the $\mathrm{Tr}\ln$. Then the action reads

$$S = \frac{1}{4J}\int_0^\beta d\tau\, \boldsymbol{\Phi}^2 - \mathrm{Tr}\ln\!\left(G_0^{-1} - W^{-1}\hat{\Sigma}W\right), \qquad G_0^{-1} = W^{-1} G_{\boldsymbol{\Phi}}^{-1} W, \qquad W(\tau) = U(\tau)\tilde{U}^{-1}(\tau) V^{-1} \tag{S4}$$

Here we introduced the matrix

$$\tilde{U}(\tau) = V^{-1} e^{\tau H \sigma_z} V. \tag{S5}$$

Since by construction $\tilde{U}(\beta) = U(\beta)$, the rotation matrix $W(\tau)$ is periodic in imaginary time, $W(\beta) = W(0)$. Therefore, the Green's function $G_0$ corresponds to the fermions in the same way as it was for $G_{\boldsymbol{\Phi}}$. We find explicitly

$$G_0^{-1} = -\partial_\tau - \hat{\epsilon} - V\tilde{U}(\tau)\partial_\tau \tilde{U}^{-1}(\tau)V^{-1} = -\partial_\tau - \hat{\epsilon} + H\sigma_z. \tag{S6}$$

Therefore, the constant $H$ plays a role of the effective Zeeman splitting. Now let us define the Green's function $G$ which satisfies the following Dyson equation:

$$G^{-1} = G_0^{-1} - \hat{\Sigma}. \tag{S7}$$

By construction $G(\tau)$ is diagonal in the spin and orbital space and is given as follows

$$G_{\alpha,\sigma}(\tau>0) = T \sum_{\varepsilon_n} \frac{e^{-i\varepsilon_n \tau}}{i\varepsilon_n - \epsilon_\alpha + H\sigma + i\gamma\,\mathrm{sgn}\,\varepsilon_n/\pi}. \tag{S8}$$

With the help of $G(\tau)$ the action can be written as

$$S = \frac{1}{4J}\int_0^\beta d\tau\,\boldsymbol{\Phi}^2 - \mathrm{Tr}\ln\Big(G^{-1} - \big(W^{-1}\hat{\Sigma}W - \hat{\Sigma}\big)\Big). \tag{S9}$$

We emphasize that the action (S9) is fully equivalent to the action defined in Eq. (1) of the main text.

### B. Wei-Norman-Kolokolov trick

For the further progress with the action (S9) we perform a change of variables. Instead of $\boldsymbol{\Phi}$ we shall use $\rho$, $\kappa$ and $\tilde{\kappa}$. These new variables are related with $\boldsymbol{\Phi}$ as follows [S1, S2, S3, S4, S5, S6]:

$$\Phi_z = \rho - 2\kappa\tilde{\kappa}, \quad \Phi_- = \tilde{\kappa}, \quad \Phi_+ = \partial_\tau \kappa + \kappa\rho - \kappa^2 \tilde{\kappa}, \tag{S10}$$

where $\Phi_\pm = (\Phi_x \pm i\Phi_y)/2$. The Jacobian of this transformation is equal to $\exp(\beta h)$ where $h = T\int_0^\beta d\tau \rho(\tau)/2$ [S6]. This transformation is supplemented by the initial condition $\kappa(0)=0$ which guarantees $U(0)=1$. In order to preserve the number of degree of freedom (field $\boldsymbol{\Phi}$ is real) we impose the following constraints: $\tilde{\kappa}=\kappa^*$ and $\rho^*=\rho$.

The $2\times 2$ matrix $U(\tau)$ can be written explicitly as

$$U(\tau) = A_\tau \begin{pmatrix} 1 & b_\tau \\ \kappa_\tau & A_\tau^{-2} + \kappa_\tau b_\tau \end{pmatrix}, \qquad A_\tau = \exp\left(\frac{1}{2}\int_0^\tau d\tau' \rho(\tau')\right), \qquad b_\tau = \int_0^\tau d\tau' \tilde{\kappa}_{\tau'} A_{\tau'}^{-2}, \tag{S11}$$

The effective Zeeman splitting $H$ is explicitly expressed in terms of the new variables:

$$\cosh\beta H = \frac{1}{2}\mathrm{tr}\,U(\beta) = \cosh\beta h + e^{\beta h}\frac{\kappa_\beta b_\beta}{2}. \tag{S12}$$

### C. Gaussian approximation

Now the partition function can be rewritten as

$$Z = \int dh\, e^{-\beta F(h)}, \qquad F(h) = \frac{h^2}{J} - h + T\,\mathrm{Tr}\ln\tilde{G} + \delta F(h), \qquad \delta F(h) = -T\ln\int D[\delta\rho, \kappa, \tilde{\kappa}]\, e^{-\delta S},$$

$$\delta S = \frac{1}{4J}\int_0^\beta d\tau\,\big(\delta\rho^2 + 4\tilde{\kappa}\partial_\tau \kappa\big) - \mathrm{Tr}\ln\Big(1 - \big(W^{-1}\hat{\Sigma}W - \hat{\Sigma} + h\sigma_z - H\sigma_z\big)\tilde{G}\Big) \tag{S13}$$

where $\delta\rho = \rho - 2h$ and the Green's function $\tilde{G}$ is given by the expression similar to Eq. (S8) with $H$ substituted by $h$. Since the Green's function $\tilde{G}$ corresponds to the free fermions in the magnetic field $h$ with the density of states $\rho(\varepsilon) = \sum_\alpha (\gamma/\pi^2)/[(\varepsilon-\epsilon_\alpha)^2 + \gamma^2/\pi^2]$:

$$\sum_\alpha \tilde{G}_{\alpha,\sigma}(i\varepsilon_n) = \int d\varepsilon \frac{\rho(\varepsilon)}{i\varepsilon_n - \varepsilon + h\sigma}, \tag{S14}$$

we find

$$T\,\mathrm{Tr}\ln\tilde{G} = -T\int d\varepsilon\,\rho(\varepsilon)\sum_\sigma \ln\Big(1 + e^{\beta(h\sigma-\varepsilon)}\Big) \approx \mathrm{const} - \frac{h^2}{\delta}. \tag{S15}$$

The quadratic dependence on $h$ holds for $\max\{h,T\} \gg \delta, \gamma$, where $\delta$ denotes the mean level spacing for single particle levels $\epsilon_\alpha$. Our aim is to write expression for $\delta S$ upto quadratic order in fields $\kappa$, $\tilde\kappa$, and $\delta\rho$. For this purpose, it is enough to expand the $\text{Tr}\ln$ up to the second order:

$$\delta S \approx \frac{1}{4J} \int_0^\beta d\tau \left( \delta\rho^2 + 4\tilde\kappa \partial_\tau \kappa \right) + \text{Tr}(H-h)\sigma_z \tilde G + \text{Tr}\left( W^{-1}\hat\Sigma W - \hat\Sigma \right)\tilde G + \frac{1}{2}\text{Tr}\left( W^{-1}\hat\Sigma W - \hat\Sigma \right)\tilde G \left( W^{-1}\hat\Sigma W - \hat\Sigma \right)\tilde G. \tag{S16}$$

Since the matrix $\hat\Sigma$ is the unit matrix in the spin space it is convenient to write down the explicit expression for the matrix $W^{-1}(\tau_1)W(\tau_2)$. We start from the expansion $U^{-1}(\tau_1)U(\tau_2) = \sum_{j=0}^3 \mathcal{P}^{(j)}(\tau_1,\tau_2)$, where

$$\mathcal{P}^{(0)} = \begin{pmatrix} A_{\tau_1}^{-1} A_{\tau_2} & 0 \\ 0 & A_{\tau_1} A_{\tau_2}^{-1} \end{pmatrix}, \quad \mathcal{P}^{(1)} = \begin{pmatrix} 0 & A_{\tau_1}^{-1} A_{\tau_2} b_{\tau_2} - b_{\tau_1} A_{\tau_1} A_{\tau_2}^{-1} \\ A_{\tau_1} A_{\tau_2}(\kappa_{\tau_2} - \kappa_{\tau_1}) & 0 \end{pmatrix},$$

$$\mathcal{P}^{(2)} = A_{\tau_1} A_{\tau_2}(\kappa_{\tau_1} - \kappa_{\tau_2}) \begin{pmatrix} b_{\tau_1} & 0 \\ 0 & -b_{\tau_2} \end{pmatrix}, \quad \mathcal{P}^{(3)} = A_{\tau_1} A_{\tau_2}(\kappa_{\tau_1} - \kappa_{\tau_2}) \begin{pmatrix} 0 & b_{\tau_1} b_{\tau_2} \\ 0 & 0 \end{pmatrix} \tag{S17}$$

As we shall see below, we need expressions for the diagonal elements of the matrix $W^{-1}(\tau_1)W(\tau_2)$ upto the second order in $\kappa, \tilde\kappa$, and $\delta\rho$, whereas for the off-diagonal components the first order is enough. All in all, we can write

$$W^{-1}(\tau_1)W(\tau_2) = \mathcal{C}^{(0)}(\tau_1,\tau_2) + \mathcal{C}^{(1)}(\tau_1,\tau_2) + \mathcal{C}^{(2)}(\tau_1,\tau_2) + \dots, \tag{S18}$$

where

$$\mathcal{C}^{(0)}(\tau_1,\tau_2) = e^{\frac{\xi_{\tau_2} - \xi_{\tau_1}}{2}\sigma_z}, \quad \mathcal{C}^{(1)}(\tau_1,\tau_2) = \begin{pmatrix} 0 & e^{-\frac{\xi_{\tau_1}+\xi_{\tau_2}}{2}}(\hat b_{\tau_2} - \hat b_{\tau_1}) \\ e^{\frac{\xi_{\tau_1}+\xi_{\tau_2}}{2}}(\hat\kappa_{\tau_2} - \hat\kappa_{\tau_1}) & 0 \end{pmatrix},$$

$$\mathcal{C}^{(2)}(\tau_1,\tau_2) = e^{\frac{\xi_{\tau_2}-\xi_{\tau_1}}{2}\sigma_z}\frac{e^{\beta h}\kappa_\beta b_\beta}{2\sinh\beta h} T(\tau_1 - \tau_2)\sigma_z + (\hat\kappa_{\tau_1} - \hat\kappa_{\tau_2}) \begin{pmatrix} e^{\frac{\xi_{\tau_2}-\xi_{\tau_1}}{2}}\hat b_{\tau_1} & 0 \\ 0 & -e^{\frac{\xi_{\tau_1}-\xi_{\tau_2}}{2}}\hat b_{\tau_2} \end{pmatrix}$$

$$+ \frac{A_\beta \kappa_\beta}{2\sinh(\beta h)} \begin{pmatrix} e^{\frac{\xi_{\tau_2}-\xi_{\tau_1}}{2}} b_{\tau_2} & 0 \\ 0 & -e^{\frac{\xi_{\tau_1}-\xi_{\tau_2}}{2}} b_{\tau_1} \end{pmatrix}. \tag{S19}$$

Here we introduced a variable $\xi$ instead of $\delta\rho$ such that $\delta\rho = \partial_\tau \xi$. We note that $\xi$ is periodic in the imaginary time, $\xi(\beta) = \xi(0)$. Also, we introduced

$$\hat b_\tau = A_\tau^2 \left( b_\tau - \frac{A_\beta b_\beta}{2\sinh(\beta h)} \right), \quad \hat\kappa_\tau = \kappa_\tau + \frac{A_\beta A_\tau^{-2} \kappa_\beta}{2\sinh(\beta h)} \tag{S20}$$

We mention that since $\kappa_0 = b_0 = 0$, the variables $\hat\kappa$ and $\hat b$ satisfy $\hat\kappa_\beta = \hat\kappa_0$ and $\hat b_\beta = \hat b_0$. Also in the course of derivation of Eq. (S19) we used the following representation for the matrix $V$:

$$V = \begin{pmatrix} v & u \\ w & v \end{pmatrix}, \quad uv = \frac{A_\beta b_\beta}{2\sinh(\beta H)}, \quad wv = -\frac{A_\beta \kappa_\beta}{2\sinh(\beta H)}, \quad v^2 = \frac{e^{\beta H} A_\beta - e^{-\beta H} A_\beta^{-1} - e^{-\beta H} A_\beta \kappa_\beta b_\beta}{2\sinh(2\beta H)},$$

$$uw = \frac{e^{-\beta H} A_\beta - e^{\beta H} A_\beta^{-1} - e^{\beta H} A_\beta \kappa_\beta b_\beta}{2\sinh(2\beta H)}. \tag{S21}$$

Let us now evaluate the terms in the right hand side of Eq. (S16). Using Eq. (S14)

$$\text{Tr}(H-h)\sigma_z \tilde G = \frac{Te^{\beta h}\kappa_\beta b_\beta}{2\sinh\beta h}\sum_\sigma \int d\varepsilon \frac{\sigma\rho(\varepsilon)}{e^{\beta(\varepsilon - h\sigma)}+1} = -4Te^{-\beta h}\sinh(\beta h)\frac{h}{\delta}\hat\kappa_\beta \hat b_\beta \tag{S22}$$

Here we approximate the density of states $\rho(\varepsilon)$ as constant equal to $\rho_d = 1/\delta$ since the mean level spacing $\delta \ll \gamma$.

Next we evaluate of the following expression:

$$\mathrm{Tr}(W^{-1}\hat{\Sigma}W - \hat{\Sigma})\tilde{G} = \int_0^\beta d\tau_1 d\tau_2 \sum_{\alpha,\sigma,\sigma'} \Big(W^{-1}_{\sigma\sigma'}(\tau_1)W_{\sigma'\sigma}(\tau_2) - 1\Big)\Sigma_\alpha(\tau_1,\tau_2)\tilde{G}_{\alpha,\sigma}(\tau_2,\tau_1)$$

$$= \int_0^\beta d\tau_1 d\tau_2\, T\sum_{\omega_n} e^{-i\omega_n(\tau_1-\tau_2)} \sum_{\sigma,\sigma'} \mathcal{X}^{(1)}_\sigma(i\omega_n)\Big(W^{-1}_{\sigma\sigma'}(\tau_1)W_{\sigma'\sigma}(\tau_2) - 1\Big)$$

$$= \int_0^\beta d\tau_1 d\tau_2\, T\sum_{\omega_n} e^{-i\omega_n(\tau_1-\tau_2)} \sum_\sigma \Big(\mathcal{X}^{(1)}_\sigma(i\omega_n) - \mathcal{X}^{(1)}_\sigma(i0)\Big) \sum_{\sigma'} \Big(W^{-1}_{\sigma\sigma'}(\tau_1)W_{\sigma'\sigma}(\tau_2) - 1\Big). \quad (S23)$$

Here we introduced

$$\mathcal{X}^{(1)}_\sigma(i\omega_n) = -\frac{i\gamma T}{\pi} \sum_{\varepsilon_n} \int d\varepsilon\, \rho(\varepsilon) \frac{\mathrm{sgn}(\varepsilon_n+\omega_n)}{i\varepsilon_n - \varepsilon + h\sigma}. \quad (S24)$$

Then, we find

$$\mathcal{X}^{(1)}_\sigma(i\omega_n) - \mathcal{X}^{(1)}_\sigma(i0) = \gamma|\omega_n|/(\pi\delta). \quad (S25)$$

Hence, we obtain

$$\mathrm{Tr}(W^{-1}\hat{\Sigma}W - \hat{\Sigma})\tilde{G} = \frac{2\gamma}{\pi\delta}\int_0^\beta d\tau_1 d\tau_2\, T\sum_{\omega_n} e^{-i\omega_n(\tau_1-\tau_2)}|\omega_n|\Big[e^{(\xi_{\tau_2}-\xi_{\tau_1})/2}\Big(1 + \frac{e^{\beta h}\kappa_\beta b_\beta}{2\sinh\beta h}T(\tau_1-\tau_2) + (\hat{\kappa}_{\tau_1}-\hat{\kappa}_{\tau_2})\hat{b}_{\tau_1}\Big) - 1\Big]$$
$$(S26)$$

Finally, expanding the exponent in powers of $\xi$, we obtain the following contribution to the Gaussian part of the action:

$$\mathrm{Tr}(W^{-1}\hat{\Sigma}W - \hat{\Sigma})\tilde{G} = -\frac{\gamma}{2\pi\delta}T\sum_{\omega_n}|\omega_n|\xi(i\omega_n)\xi(-i\omega_n) - \frac{2\gamma}{\pi\delta}T\sum_{\omega_n}|\omega_n|\hat{\kappa}(i\omega_n)\hat{b}(-i\omega_n). \quad (S27)$$

Next we consider the last term in the right hand side of Eq. (S16):

$$\frac{1}{2}\mathrm{Tr}(W^{-1}\hat{\Sigma}W - \hat{\Sigma})\tilde{G}(W^{-1}\hat{\Sigma}W - \hat{\Sigma})\tilde{G} = T^2 \sum_{\varepsilon_n,\omega_n} \mathcal{C}^{(1)}_{\sigma\sigma'}(i\omega_n)\mathcal{C}^{(1)}_{\sigma'\sigma}(-i\omega_n) \sum_{\alpha,\sigma,\sigma'} \big[\Sigma(i\varepsilon_n) - \Sigma(i\varepsilon_n+i\omega_n)\big]$$

$$\times \Sigma(i\varepsilon_n+i\omega_n)\tilde{G}_{\alpha,\sigma'}(i\varepsilon_n)\tilde{G}_{\alpha,\sigma}(i\varepsilon_n+i\omega_n) = \frac{2\gamma^2}{\pi^2\delta}T\sum_{\omega_n,\sigma,\sigma'} \frac{i\omega_n}{i\omega_n + h(\sigma-\sigma') + 2i\gamma\,\mathrm{sgn}\,\omega_n/\pi} \mathcal{C}^{(1)}_{\sigma\sigma'}(i\omega_n)\mathcal{C}^{(1)}_{\sigma'\sigma}(-i\omega_n)$$

$$= \frac{\gamma^2}{\pi^2\delta}T\sum_{\omega_n} \frac{i\omega_n}{i\omega_n + 2i\gamma\,\mathrm{sgn}\,\omega_n/\pi}\xi(i\omega_n)\xi(-i\omega_n) + \frac{4\gamma^2}{\pi^2\delta}T\sum_{\omega_n} \frac{i\omega_n}{i\omega_n - 2h + 2i\gamma\,\mathrm{sgn}\,\omega_n/\pi}\hat{\kappa}(i\omega_n)\hat{b}(-i\omega_n) \quad (S28)$$

Here we used the following simplification. In order to derive the contribution to the Gaussian part of the action we can neglect $\xi$ in the expression for $\mathcal{C}^{(1)}(\tau_1,\tau_2)$, Eq. (S19). Then we expressed it as follows

$$\mathcal{C}^{(1)}(\tau_1,\tau_2) \approx \mathcal{C}^{(1)}(\tau_2) - \mathcal{C}^{(1)}(\tau_1), \qquad \mathcal{C}^{(1)}(\tau) = \begin{pmatrix} 0 & \hat{b}_\tau \\ \hat{\kappa}_\tau & 0 \end{pmatrix}. \quad (S29)$$

Next, we perform the following transformation:

$$\int_0^\beta d\tau\, \tilde{\kappa}\partial_\tau \kappa = \int_0^\beta d\tau\, \big[\partial_\tau \hat{b}_\tau - 2h\hat{b}_\tau\big]\big[\partial_\tau \hat{\kappa}_\tau + 2he^{-2h\tau}\hat{\kappa}_\beta\big] = T\sum_{\omega_n}\omega_n\big[\omega_n + i2h\big]\hat{\kappa}(i\omega_n)\hat{b}(-i\omega_n) - 4he^{-\beta h}\sinh(\beta h)\hat{\kappa}_\beta \hat{b}_\beta. \quad (S30)$$

All in all, we obtain

$$\delta S = -\left(\frac{1}{\delta}+\frac{1}{J}\right)4he^{-\beta h}\sinh(\beta h)\hat{\kappa}_\beta \hat{b}_\beta + \frac{T}{4J}\sum_{\omega_n}\omega_n^2\left[1-\frac{J}{\delta}\frac{2i\gamma\,\text{sgn}\,\omega_n/\pi}{i\omega_n+2i\gamma\,\text{sgn}\,\omega_n/\pi}\right]\xi(i\omega_n)\xi(-i\omega_n)$$
$$+\frac{T}{J}\sum_{\omega_n}\omega_n[\omega_n+i2h]\left[1-\frac{J}{\delta}\frac{2i\gamma\,\text{sgn}\,\omega_n/\pi}{i\omega_n-2h+2i\gamma\,\text{sgn}\,\omega_n/\pi}\right]\hat{\kappa}(i\omega_n)\hat{b}(-i\omega_n). \tag{S31}$$

Within the Gaussian approximation we have the relation $\tilde{\kappa}_\tau = \partial_\tau \hat{b}_\tau - 2h\hat{b}_\tau$. This implies that $\tilde{\kappa}_\beta = \tilde{\kappa}_0$. Since $\tilde{\kappa}_\tau = \kappa_\tau^*$ and $\kappa_0 = 0$ we find that the relation $\tilde{\kappa}_\beta = \tilde{\kappa}_0 = \kappa_\beta = \kappa_0 = 0$ holds within the Gaussian approximation. We emphasize that there is no periodic boundary conditions for the fields $\kappa_\tau$ and $\tilde{\kappa}_\tau$ in general.

Finally, we find

$$\delta S = \frac{T}{2J}\sum_{\omega_n>0}\omega_n^2\left[\frac{\omega_n+2a\gamma/\pi}{\omega_n+2\gamma/\pi}\right]|\xi(i\omega_n)|^2 + \frac{T}{J}\sum_{\omega_n}(-i\omega_n)\frac{i\omega_n-2h+2ia\gamma\,\text{sgn}\,\omega_n/\pi}{i\omega_n-2h+2i\gamma\,\text{sgn}\,\omega_n/\pi}|\kappa(i\omega_n)|^2, \tag{S32}$$

where $a = 1 - J/\delta$. Now we can perform Gaussian integration over $\delta\rho$, $\kappa$, and $\tilde{\kappa}$. Then, we find

$$F(h) = \frac{h^2}{J_*} - h + 2T\,\text{Re}\,\ln\frac{\Gamma\left(1+\frac{ih}{\pi T}+\frac{\gamma}{\pi^2 T}\right)}{\Gamma\left(1+\frac{a\gamma}{\pi^2 T}+\frac{ih}{\pi T}\right)\Gamma\left(1+\frac{\gamma}{\pi^2 T}\right)}. \tag{S33}$$

We note that the free energy (S33) coincides with Eq. (3) of the main text for $a = 0$. As follows from the inequalities (4) of the main text for $\gamma \ll J_*^2/J$ the Gaussian approximation is valid for $|h|, T \gg a\gamma$. Since we are interested in the vicinity of the QCP at $\gamma_c = J_*$ we can work with the free energy with $a = 0$.

### D. Limitations for the Gaussian approximation

As we have pointed out above, the computation of the Gaussian part of the action for the fields $\xi$, $\kappa$, and $\tilde{\kappa}$ requires the expansion of $\exp\xi$ in powers of $\xi$. Therefore, the Gaussian approximation is legitimate under assumption $\langle\xi^2(\tau)\rangle \ll 1$. Using Eq. (S32), we find

$$\langle\xi^2(\tau)\rangle = 2T^2\sum_{\omega_n>0}\langle|\xi(i\omega_n)|^2\rangle = 4JT\sum_{\omega>0}\frac{1}{\omega_n^2}\frac{\omega_n+2\gamma/\pi}{\omega_n+2a\gamma/\pi} = \frac{J}{\pi^2 Ta}\left[\frac{\pi^2}{6}-\frac{1-a}{a}\frac{\pi^2 T}{\gamma}\left(\psi\left(1+\frac{a\gamma}{\pi^2 T}\right)-\psi(1)\right)\right]. \tag{S34}$$

Then we see that the condition $\langle\xi^2(\tau)\rangle \ll 1$ is equivalent to the condition $T \gg \max\{J, \min\{J_*, \sqrt{J\gamma}\}\}$, cf. Eq. (4) from the main text.

Also, in derivation of the Gaussian part of the action we neglect the contribution due to $\mathcal{P}^{(3)}$. Then, within the second order expansion of $\text{tr}\ln$ we find the additional term which involves two fields $\hat{\kappa}$ and two fields $\hat{b}$. Such term produces the following correction to the Gaussian action

$$\frac{2\gamma^2}{\pi^2\delta}T^3\sum_{\omega_n,\Omega_n}\min\{|\omega_n|,|\Omega_n|\}\left[\frac{i\,\text{sgn}\,\Omega_n}{i\Omega_n-2h+2i\gamma\,\text{sgn}\,\Omega_n/\pi}+\frac{i\,\text{sgn}\,\omega_n}{i\omega_n-2h+2i\gamma\,\text{sgn}\,\omega_n/\pi}\right]\langle\hat{\kappa}(i\Omega_n)\hat{b}(-i\Omega_n)\rangle\hat{\kappa}(i\omega_n)\hat{b}(-i\omega_n)$$
$$= \frac{2\gamma^2 J}{\pi^2\delta}T^2\sum_{\omega_n,\Omega_n}\min\{|\omega_n|,|\Omega_n|\}\left[\frac{i\,\text{sgn}\,\Omega_n}{i\Omega_n-2h+2i\gamma\,\text{sgn}\,\Omega_n/\pi}+\frac{i\,\text{sgn}\,\omega_n}{i\omega_n-2h+2i\gamma\,\text{sgn}\,\omega_n/\pi}\right]$$
$$\times\frac{i\Omega_n-2h+2i\gamma\,\text{sgn}\,\Omega_n/\pi}{(-i\Omega_n)(i\Omega_n-2h)(i\Omega_n-2h+2ia\gamma\,\text{sgn}\,\Omega_n/\pi)}\hat{\kappa}(i\omega_n)\hat{b}(-i\omega_n) \tag{S35}$$

Here we used Eq. (S32). Performing summation over $\Omega_n$, we obtain

$$\frac{2\gamma^2 J}{\pi^3 a\delta} T \sum_{\omega_n} \hat\kappa(i\omega_n)\hat b(-i\omega_n) \Bigg\{ \frac{i\,\mathrm{sgn}\,\omega_n}{i\omega_n - 2h + 2i\gamma\,\mathrm{sgn}\,\omega_n/\pi} \mathrm{Re}\Bigg[(1-a)\psi\left(1+\frac{ih}{\pi T}+\frac{ia\gamma}{\pi^2 T}\right) + \left(1-\frac{i|\omega_n|}{2h}\right)\psi\left(1+\frac{|\omega_n|}{2\pi T}+\frac{ih}{\pi T}\right)$$

$$-(1-a)\frac{i|\omega_n|-2h+2ia\gamma/\pi}{-2h+2ia\gamma/\pi}\psi\left(1+\frac{|\omega_n|}{2\pi T}+\frac{ih}{\pi T}+\frac{ia\gamma}{\pi^2 T}\right) - \psi\left(1+\frac{ih}{\pi T}\right) + \frac{ia|\omega_n|}{2h}\frac{-h+i\gamma/\pi}{-h+ia\gamma/\pi}\psi\left(1+\frac{|\omega_n|}{2\pi T}\right)\Bigg]$$

$$+\frac{\pi}{2\gamma}\mathrm{Re}\Bigg[\frac{i|\omega_n|-2h+2ia\gamma/\pi}{2h-2ia\gamma/\pi}\psi\left(1+\frac{|\omega_n|}{2\pi T}+\frac{ih}{\pi T}+\frac{ia\gamma}{\pi^2 T}\right) + \left(1-\frac{i|\omega_n|}{2h}\right)\psi\left(1+\frac{|\omega_n|}{2\pi T}+\frac{ih}{\pi T}\right)$$

$$+\psi\left(1+\frac{ih}{\pi T}+\frac{ia\gamma}{\pi^2 T}\right) - \psi\left(1+\frac{ih}{\pi T}\right) + \frac{|\omega_n|}{2h}\frac{a\gamma/\pi}{h-ia\gamma/\pi}\psi\left(1+\frac{|\omega_n|}{2\pi T}\right)\Bigg]\Bigg\} \quad (S36)$$

This contribution is small in comparison with the last term in Eq. (S28) for all values of $|\omega_n|$ provided the inequalities $|h|, T \gg \max\{J, \min\{J_*, \sqrt{J\gamma}\}\}$, cf. Eq. (4) of the main text, are fulfilled.

As one can check the other higher order corrections are small under conditions (4) of the main text.

## S.II. SPIN SUSCEPTIBILITY

The spin susceptibility of electrons on the QD can be computed as [S7]

$$\chi = \frac{T}{12 J^2}\left\langle\left|\int_0^\beta d\tau\,\mathbf{\Phi}\right|^2\right\rangle - \frac{1}{2J}. \quad (S37)$$

Substituting the representation (S10), we obtain

$$\left|\int_0^\beta d\tau\,\mathbf{\Phi}\right|^2 = 4\beta^2 h^2 + 4\kappa_\beta \int_0^\beta d\tau\,\tilde\kappa_\tau + 4\int_0^\beta d\tau_1 d\tau_2\, \kappa_{\tau_1}\tilde\kappa_{\tau_2}(\delta\rho_{\tau_1}-\delta\rho_{\tau_2}) + 4\int_0^\beta d\tau_1 d\tau_2\, \kappa_{\tau_1}\tilde\kappa_{\tau_1}\tilde\kappa_{\tau_2}(\kappa_{\tau_2}-\kappa_{\tau_1}), \quad (S38)$$

Performing averaging with respect to the Gaussian part of the action, $\delta S$, we find

$$\left|\int_0^\beta d\tau\,\mathbf{\Phi}\right|^2 = 4\beta^2 h^2 + 16 J^2\Bigg[\mathrm{Im}\sum_{\omega_n>0}\frac{1}{\omega_n}\frac{\omega_n+2ih+2\gamma}{\omega_n+2ih+2a\gamma}\Bigg]^2 - 8J^2\mathrm{Re}\sum_{\omega_n>0}\frac{1}{\omega_n^2}\Bigg[\frac{\omega_n+2ih+2\gamma}{\omega_n+2ih+2a\gamma}\Bigg]^2$$

$$= 4\beta^2 h^2 + \frac{4\beta^2 J^4 \gamma^2}{\pi^2\delta^2(h^2+a^2\gamma^2)^2}\Bigg\{h\Big[\psi(1)-\mathrm{Re}\,\psi\left(1+\frac{a\gamma+ih}{\pi T}\right)\Big] + a\gamma\,\mathrm{Im}\,\psi\left(1+\frac{a\gamma+ih}{\pi T}\right)\Bigg\}^2$$

$$-\frac{2\beta^2 J^2}{\pi^2}\mathrm{Re}\Bigg\{\frac{\pi^2}{6}\frac{(\gamma+ih)^2}{(a\gamma+ih)^2} + \frac{2\pi T J}{\delta}\frac{\gamma(\gamma+ih)}{(a\gamma+ih)^3}\Big[\psi(1)-\psi\left(1+\frac{a\gamma+ih}{\pi T}\right)\Big]$$

$$+\frac{J^2}{\delta^2}\frac{\gamma^2}{(a\gamma+ih)^2}\psi'\left(1+\frac{a\gamma+ih}{\pi T}\right)\Bigg\} \quad (S39)$$

Under the conditions $|h|, T \gg \max\{J, \min\{J_*, \sqrt{J\gamma}\}\}$, cf. Eq. (4) of the main text, we find that

$$\left|\int_0^\beta d\tau\,\mathbf{\Phi}\right|^2 \approx 4\beta^2 h^2. \quad (S40)$$

Hence we obtain the result (5) of the main text.

## S.III. THE EFFECT OF TUNNELING NEAR AND AWAY FROM THE VICINITY OF THE MACROSCOPIC STONER TRANSITION

As it was discussed in the Introduction (see the main text), in the absence of tunneling there is a finite region of $J < \delta$ at which the QD in the ground state has non zero total spin. The states with difference values of the total spin

are separated by the first order quantum phase transitions. These transitions can be seen from the exact partition function for the isolated quantum dot [S7, S7]:

$$Z^{(0)} \propto \int dh\, h \sinh(\beta h) \prod_\alpha \left(1+e^{-\beta(\epsilon_\alpha-h)}\right)\left(1+e^{-\beta(\epsilon_\alpha+h)}\right) e^{-\beta h^2/J}$$
$$\propto \int dh\, h \sinh(\beta h) \sum_{n_\uparrow,n_\downarrow} Z_{n_\uparrow} Z_{n_\downarrow} e^{-\beta h(n_\uparrow-n_\downarrow)-\beta h^2/J}, \tag{S41}$$

where we use the Darwin-Fowler integral:

$$Z_n = \int_0^{2\pi} \frac{d\theta}{2\pi} e^{-i\theta n} \prod_\alpha \left(1+e^{i\theta-\beta\epsilon_\alpha}\right). \tag{S42}$$

Introducing the integer or half-integer value $m=(n_\uparrow-n_\downarrow)/2$, we can rewrite the part of the partition function that depends on $h$ as follows

$$Z^{(0)} \propto \int dh\, h \sinh(\beta h) \sum_m e^{\beta\delta m^2+2\beta hm-\beta h^2/J}. \tag{S43}$$

We note that integration over $m$ instead of summations produces the parabolic dependence on $h$: $\beta h^2/\delta$. In this way, one obtains the free energy given by Eq. (3) of the main text with $\gamma=0$. The existence of the first order QPTs is due to discreteness of $m$ which makes piecewise-linear function instead of parabola.

In the presence of a weak tunneling, $\gamma \ll \delta$, the partition function can be still computed exactly to the lowest order in $\gamma$ by the method described in Refs. [S7, S7]. For $|h| \gg T, \delta$ one finds

$$Z \propto \int dh\, h \sinh(\beta h) \sum_m e^{\beta\delta m^2+2\beta hm-\beta h^2/J-2(\beta\gamma/\pi^2)\ln(\beta|h|)}. \tag{S44}$$

We note that the term proportional to $\gamma$ is exactly the same as one can find from the result (3) of the main text. The reason is that for $\gamma \ll J < \delta$ the Gaussian approximation is justified for $|h| \gg T \gg \delta$. In order to find the energy of the state with the given $m$ one needs to integrate over $h$. For large non-negative values of $m$, we can do it within the saddle point method. There is the saddle point

$$h \approx J(m+1/2) - \frac{\gamma}{\pi^2(m+1/2)}. \tag{S45}$$

Hence the energy of the state with spin $S \equiv m \geq 0$ is given as

$$E(S) = \delta S^2 - J(S+1/2)^2 + \frac{2\gamma}{\pi^2}\ln(S+1/2) \tag{S46}$$

We note that above we did not take into account the charging energy. The role of the charging energy is to fix the total number of electron in the regime of the Coulomb valley. This means that then $S$ can be changed only by $\Delta S = \pm 1$. In the absence of the charging energy or in the case of the Coulomb peak the total number of electrons is not fixed such that the total spin can be changed by $\Delta S = \pm 1/2$. Using Eq. (S46), one can check that in both cases of $\Delta S = \pm 1/2, \pm 1$ a weak tunneling, $\gamma \ll \delta$, does not destroy the first order quantum phase transition between the ground states with different values of $S$.

For $\gamma \gg \delta$, the most crucial affect of tunneling is broadening of the single-particle levels on the quantum dot. Since for $\gamma \gg \delta$ the density of states $\rho(\varepsilon)$ is almost the constant with exponential accuracy, we obtain, cf. Eq. (S15),

$$\det \tilde{G} = \exp \int d\varepsilon \rho(\varepsilon) \left(1+e^{-\beta(\varepsilon-h)}\right)\left(1+e^{-\beta(\varepsilon+h)}\right) \propto e^{\beta h^2/\delta} \tag{S47}$$

Therefore, due to the broadening of the single particle levels this contribution to the free energy is exactly parabola. In other words, at $\gamma \gg \delta$ the discreteness of the total spin is completely smeared. This implies the absence of the first order quantum phase transitions at $\gamma \gg \delta$. Therefore, it is natural to expect that the first order quantum phase transitions survive until $\gamma \sim \delta$ only.

---

[S1] J. Wei and E. Norman, *Lie algebraic solution of linear differential equations*, J. Math. Phys. **4**, 575 (1963).




\begin{thebibliography}{100}
\bibitem{footnote0} In non-ferromagnetic materials the corresponding  ferromagnetic exchange is bounded from above, $J<\delta$. This bound is related with the fact that for  a QD of size much larger than the Fermi wave length $J$ can be estimated as $J/\delta = -F_0^\sigma$. Here $F_0^\sigma$ is the Fermi-liquid interaction parameter of the corresponding bulk material. The Stoner transition occurs at $F_0^\sigma=-1$.
\bibitem{footnote4} We note $J$ and $\delta$ scales identically with the system size reflecting the non-extensive behavior of the non-zero spin in the ground state.
\end{thebibliography}
\end{document}